\documentclass[pra,10pt,aps,twocolumn,floatfix]{revtex4-1} 

\usepackage{amsmath,amsfonts,amssymb,graphicx,color,bbm}

\usepackage{hyperref}

\DeclareMathOperator{\tr}{tr}

\begin{document}

\bibliographystyle{apsrev}
\newtheorem{theorem}{Theorem}
\newtheorem{corollary}{Corollary}
\newtheorem{definition}{Definition}
\newtheorem{proposition}{Proposition}
\newtheorem{lemma}{Lemma}
\newcommand{\proofend}{\hfill\fbox\\\medskip }
\newcommand{\proof}[1]{\textbf{Proof.} #1 $\proofend$}
\newcommand{\nn}{{\mathbbm{N}}}
\newcommand{\rr}{{\mathbbm{R}}}
\newcommand{\cc}{{\mathbbm{C}}}
\newcommand{\mbp}{\ensuremath{\spadesuit}}
\newcommand{\je}{\ensuremath{\heartsuit}}
\newcommand{\jd}{\ensuremath{\clubsuit}}
\newcommand{\id}{{\mathbbm{1}}}
\renewcommand{\vec}[1]{\boldsymbol{#1}}
\newcommand{\me}{\mathrm{e}}
\newcommand{\mi}{\mathrm{i}}
\newcommand{\md}{\mathrm{d}}
\newcommand{\sg}{\text{sgn}}

\def\>{\rangle}
\def\<{\langle}
\def\({\left(}
\def\){\right)}

\newcommand{\ket}[1]{|#1\>}
\newcommand{\bra}[1]{\<#1|}
\newcommand{\braket}[2]{\<#1|#2\>}
\newcommand{\ketbra}[2]{|#1\>\!\<#2|}
\newcommand{\proj}[1]{|#1\>\!\<#1|}
\newcommand{\avg}[1]{\< #1 \>}

\renewcommand{\tensor}{\otimes}

\delimitershortfall=-2pt

\title{Efficient quantum state tomography}

\author{Marcus~Cramer}
\affiliation{Institut f\"ur Theoretische Physik,
Albert-Einstein Allee 11, Universit\"at Ulm, 89069 Ulm, Germany}

\author{ Martin B.~Plenio}
\affiliation{Institut f\"ur Theoretische Physik,
Albert-Einstein Allee 11, Universit\"at Ulm, 89069 Ulm, Germany}

\author{Steven T. Flammia}
\affiliation{Perimeter Institute for Theoretical Physics, Waterloo, Ontario, N2L 2Y5 Canada}

\author{David Gross}
\affiliation{Institute for Theoretical Physics, Leibniz University Hannover, 30167 Hannover, Germany}

\author{Stephen D. Bartlett}
\affiliation{School of Physics, The University of Sydney, Sydney, New South Wales 2006, Australia}

\author{Rolando Somma}
\affiliation{Perimeter Institute for Theoretical Physics, Waterloo, Ontario, N2L 2Y5 Canada}

\author{Olivier Landon-Cardinal}
\affiliation{D\'epartement de Physique, Universit\'e de Sherbrooke, Sherbrooke, Qu\'ebec, Canada}

\author{Yi-Kai Liu}
\affiliation{Institute for Quantum Information, California Institute of Technology, Pasadena, CA, USA}

\author{David Poulin}
\affiliation{D\'epartement de Physique, Universit\'e de Sherbrooke, Sherbrooke, Qu\'ebec, Canada}

\begin{abstract}
Quantum state tomography \cite{Vogel R 89}, the ability to deduce the
state of a quantum system from measured data,
is the gold standard for verification and benchmarking of quantum devices.
It has been realized in systems with few components
\cite{Smithey 93, Haffner 05, Leibfried 05, James KMW 01, Dunn, Lvovsky 09},
but for larger systems it becomes infeasible because the number
of quantum measurements and the amount of computation required to process
them grows exponentially in the system size. Here we show that we can do
exponentially better than direct state tomography for a wide range of 
quantum states, in particular those that are well approximated by a matrix 
product state ansatz. We present two schemes for tomography in 1-D quantum 
systems and touch on generalizations. One scheme requires unitary operations on a
constant number of subsystems, while the other requires only local measurements
together with more elaborate post-processing. Both schemes rely only on a linear
number of experimental operations and classical postprocessing that is polynomial
in the system size. A further strength of the methods is that the accuracy of the reconstructed states can be rigorously certified without any \textit{a priori} assumptions.
\end{abstract}

\maketitle

\date{\today}

\section{Introduction}
One of the principal features distinguishing classical from
quantum many-body systems is that quantum systems require exponentially many parameters in the system size to fully specify the state, compared to only linearly many for classical systems. Put to use constructively, the exponential complexity
enables the construction of information processing devices
fundamentally superior to any classical device. At the same time,
however, this ``curse of dimensionality'' makes engineering
tasks---such as verifying that the quantum processing device
functions as intended---a daunting challenge.

The full determination of the quantum state of a system, known as
quantum state tomography \cite{Vogel R 89}, has already been
achieved by measuring a complete set of observables whose
expectation values determine the quantum state \cite{Smithey 93,
Haffner 05, Leibfried 05, James KMW 01, Dunn, Lvovsky 09}.  As it is typically
formulated \cite{Paris2004}, simply to output an estimate for a
generic state would take exponential time in the system size $N$, given
that there are an exponential number of coefficients in a generic
state's description.  This is but one of several inefficiencies. Most
quantum states have exponentially small amplitudes in almost every
basis, so to distinguish any one of those amplitudes from zero, one must take
an exponential number of samples.  Assuming one were able to collect all
the data from an informationally complete measurement, one is left
with the intractable computational task of inverting the measured
frequencies to find an estimate of the state.

However, the traditional representation of quantum states
is in a sense too general. Indeed, states which occur in many
practical situations are specified by a small number of parameters. An
efficient description could be a practical preparation scheme which
outputs the state; or, in the case of thermodynamical equilibrium
states, a local Hamiltonian and a temperature. This insight is not
new: researchers in many-body physics and quantum information theory
have found many classes of states which are described by a number of
parameters scaling polynomially in $N$ \cite{FannesNW 92, Rommer 95,Perez-Garcia2007a, Vidal2007,Rizzi} and which closely approximate
the kind of states found in physical systems
\cite{Hastings2006, Hastings2007a}.
However, the question of whether these restricted classes can be put
to use in the context of tomography has remained largely open.

In this work, we address the above challenge. The physical system we
have in mind is one where the constituents are arranged in a
one-dimensional configuration (e.g., ions in a linear trap
\cite{Haffner 05}). But the methods that we are presenting here can be
generalized to higher-dimensional arrangements such as those realized
in optical lattices \cite{Zwerger}.  It is highly plausible that in such a 
setting, correlations
between neighboring particles are much more pronounced (due to direct
interaction) than correlations between distant systems (mediated e.g.\
by global fluctuations of control fields). An efficiently describable
class of states anticipating exactly this behavior has long been
studied under the names of finitely correlated states (FCS) or matrix
product states (MPS) \cite{FannesNW 92,Perez-Garcia2007a}. 
Importantly, restricting to this class is not a limitation since
\emph{every} state may be written as a MPS with a suitable, albeit
possibly large, matrix dimension.  Since many states that are relevant for quantum 
information processing or quantum many-body physics 
have a small (independent of $N$) bond dimension, our methods are 
directly applicable to such states; examples include, 
but are not limited to, ground and thermal states, the 
GHZ, W, cluster, and AKLT states, the latter two being universal resources states for quantum computing.

Given that standard tomography is no longer feasible in a range of 
recent and upcoming experiments involving large numbers of qubits, 
our results represent a significant advance in the ability to verify 
and quantitatively and efficiently benchmark systems 
of experimental importance.

\section{Results}

In the following we present two schemes for identifying systems that are well approximated by an MPS, initially focusing on pure states for simplicity. We 
will view each system as consisting of a linear chain of $N$ qudits, each 
having dimension $d$. Both schemes require the measurement of linearly (in 
the system size $N$) many local observables within finite accuracy, 
polynomial classical post-processing of the data and can certify
the accuracy of the reconstructed state without making any technical assumptions
about the state in the laboratory. The first scheme requires unitary control 
and local measurements while the second scheme removes the need for unitary 
control at the cost of more elaborate post-processing.

\subsection{Scheme based on unitary transformations}

The key idea of the method consists of a sequential procedure to
\emph{disentangle} the left hand side of the chain from the right 
hand side, using a sequence of unitary operations with small interaction 
length independent of $N$.  The result will be a product 
state and a sequence of local unitary operations from which to construct the
original state.

\begin{figure}
\includegraphics[width=0.9\columnwidth]{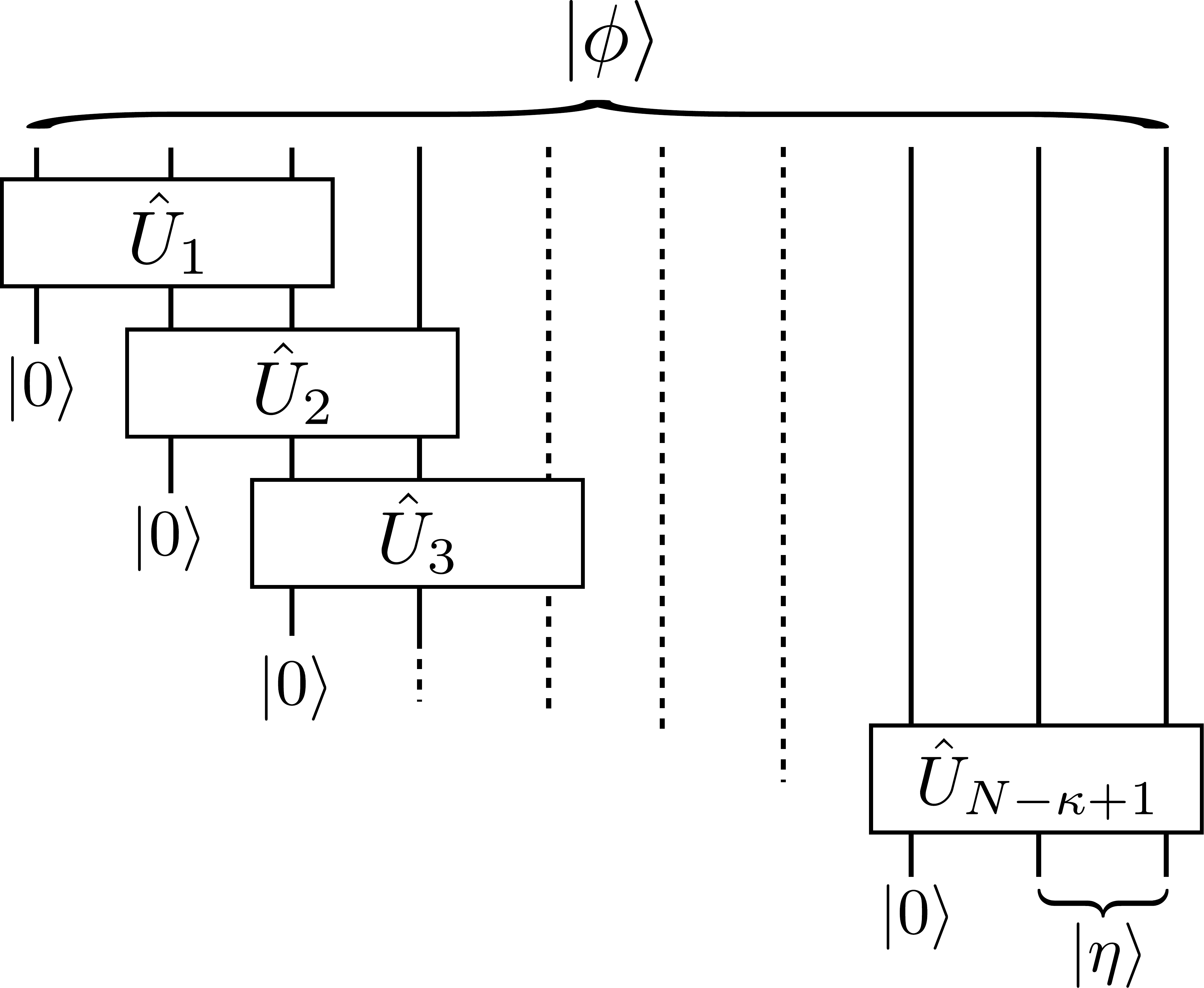}
\caption{\label{fig_circuit} 
Quantum circuit that transforms 
$\ket{\phi}$ into $\ket{0}^{\otimes N-\kappa+1}\otimes \ket{\eta}$
with $\kappa=3$. The unitaries $\hat{U}_i$ successively disentangle the particles
and the state $\ket{\eta}$ on the last sites acts as a boundary condition to 
determine the MPS description of $\ket{\phi}$.}
\end{figure}

Suppose the ideal state in the laboratory is $\hat\varrho = |\phi\rangle\langle\phi|$, which we assume for clarity is a MPS of given 
bond dimension. This implies that the rank of reductions to one part of a 
bipartite (left vs.\ right) split of the chain is bounded by 
a constant $R$. The protocol starts by estimating, through standard state
tomography, the reduced density matrix of the first $\kappa=\lceil\log_d(R)\rceil+1$ sites, $\hat{\sigma} \approx \text{tr}_{\kappa+1,\dots,N}[\hat\varrho]$. This reduced 
density matrix has the eigendecomposition $\hat{\sigma}=\sum_{r=1}^R\sigma_r|\phi_r\rangle\langle\phi_r|$ where
the rank $R \leq d^{\kappa-1}$. 
Hence there exists a density matrix with one fewer qudit that has the same rank $R$ and eigenvalues $\sigma_r$ as $\hat{\sigma}$.
Therefore we can disentangle the first site in $\hat{\sigma}$ with the following 
unitary acting on the first $\kappa$ sites:
\begin{equation}
    \hat{U}=\sum_{s=0}^{d-1}\sum_{s^\prime=0}^{d^{\kappa -1}-1}\ket{s}_1\otimes \ket{s^\prime}_{2,\dots,\kappa }\bra{\phi_{sd^{\kappa -1}+s^\prime+1}}_{1,\dots,\kappa},
\end{equation}
where $\ket{\phi_1},\ldots,\ket{\phi_R}$ have been extended in some arbitrary 
way to get a complete basis for sites $1,\ldots,\kappa$. Applying $\hat{U}$ 
produces the state $\hat{U}|\phi\rangle=|0\rangle_1\otimes|v\rangle_{2,\dots,N}$,
where $|v\rangle$ is some pure state on sites $2,\dots, N$. Hence $\hat{U}$ disentangles the first qudit from all the others. Now, set aside this first 
qudit, look at sites $2,\dots,\kappa +1$, and repeat the above process as 
indicated on Fig.~\ref{fig_circuit}. In this way, one obtains a sequence of
unitaries $\hat{U}_1,\dots,\hat{U}_{N-\kappa +1}$, where each $\hat{U}_i$
acts on sites $i,\dots,i+\kappa -1$. This sequence transforms $|\phi\rangle$ into $\hat{U}_{N-\kappa +1}\cdots\hat{U}_1|\phi\rangle=|0\rangle^{\otimes N-\kappa +1}\otimes |\eta\rangle$, where $|\eta\rangle$ is some pure state on the last
$\kappa -1$ sites. 

In summary, this scheme infers the quantum circuit used to prepare an MPS \cite{schoen05}. The MPS decomposition of $|\phi\rangle$ can then be obtained 
readily from the $U_i$ and $\ket\eta$~\cite{Vidal2003a}.
If the state in the laboratory $\hat{\varrho}$ is arbitrary, then 
the reduced density matrices $\hat{\sigma}$ will generally have full rank. Hence in each
step we will need to truncate the $\kappa$ qudit state $\hat{\sigma}$ to a rank $R$
subspace with $R<d^{\kappa-1}$.  Then the above method will produce an MPS approximation to $\hat{\varrho}$.  The accuracy of this estimate can be 
certified, without any assumptions on the state, by keeping track of the effects 
of truncating each of the reduced states $\hat{\sigma}$. As shown in the 
Methods, errors of magnitude $\epsilon$ due to finite measurement precision or truncation error (as measured by the weight of the truncated space) 
accumulate at most linearly with the number of sites, and 
can be evaluated directly from the data, leading to an estimate of accuracy $N\epsilon$.

The present scheme requires unitary control of $\kappa$ neighboring qudits, 
which is challenging to implement in many current experimental settings. We 
now present a second scheme that avoids unitary control and requires only 
local measurements.

\subsection{Scheme based on local measurements: Certification of estimated state}\label{S:certification}

Consider again the state $\hat{\varrho}$ in the laboratory and suppose 
that a tomographically-complete set of local measurements on all groups 
of $k$ neighbouring qubits has been performed. Further suppose that 
sufficient data has been taken to yield estimates 
$\hat{\sigma}_i$ of the local reductions
$\hat{\varrho}_i=\text{tr}_{1,\dots,i;i+k+1,\dots,N}[\hat{\varrho}]$
such that
\begin{equation}
\|\hat{\varrho}_i-\hat{\sigma}_i\|_{\text{tr}}\le\epsilon_i \, .
\end{equation}
Determining these approximate reduced density operators
$\hat{\sigma}_i$ completes the experimental work. In the remainder we 
describe the classical post-processing that will result in a MPS 
estimate $|\psi\rangle$ to $\hat{\varrho}$ and a lower bound to the 
fidelity $\langle\psi|\hat{\varrho}|\psi\rangle$ \cite{Gilchrist}
that does {\em not} require assumptions on the nature of the state 
$\hat{\varrho}$.

We start with the latter, since the following easy calculation yields 
the fidelity bound we have in mind and hints at the MPS estimate we 
are after. Suppose, for the sake of the argument, that $|\psi\rangle$ 
is the unique ground state (with energy zero) of a local Hamiltonian $\hat{H}=\sum_i\hat{h}_i$, where the $\hat{h}_i$ is a projection operator 
acting only on sites $i+1,\dots,i+k$ (as it turns out,
generic MPS are of this type). 
Then, expanding in the eigenbasis $\hat{H}=\sum_{n=0}^{2^N-1} E_n|E_n\rangle\langle E_n|$,
\begin{equation}
  \label{eq:localerror}
    \text{tr}[\hat{H}\hat{\varrho}]\ge \Delta E\sum_{n>0} \langle E_n|\hat{\varrho}|E_n\rangle
    = \Delta E(1-\langle\psi|\hat{\varrho}|\psi\rangle),
\end{equation}
where we denoted by $\Delta E$ the energy gap above the ground state 
$|\psi\rangle$. Hence, we have the 
fidelity bound
\begin{align}\label{E:fidelitybound}
    \langle\psi|\hat{\varrho}|\psi\rangle 
    \ge 1-\frac{\sum_i\text{tr}[h_i \hat{\rho}]}{\Delta E}
    \ge 1-\frac1{\Delta E}\big(\sum_i\text{tr}[\hat{h}_i \hat{\sigma}_i] + \epsilon_i\big).
\end{align}
In other words, the Hamiltonian acts as a \emph{witness} for its ground
state $|\psi\rangle$. This bound is tight: suppose the experimental
estimates $\hat{\sigma}_i$ and the reductions of the state in the
laboratory $\hat{\varrho}_i$ coincide, that is $\epsilon_i=0$. If, in
addition, the reductions of $|\psi\rangle$ match the $\hat{\sigma}_i$
then, as $|\psi\rangle$ was assumed to be the unique ground state with
energy zero of $\sum_i\hat{h}_i$, we have $\sum_i\text{tr}[\hat{h}_i\hat{\sigma}_i]=0$, i.e., $\langle\psi|\hat{\varrho}|\psi\rangle=1$.

The goal is now clear: find a local
gapped Hamiltonian $\hat{H}$ such that the reductions of its ground state
are close to the $\hat{\sigma}_i$. \emph{A priori} it is unclear whether such
a convenient witness always exists and how it could be found efficiently. 
However, using formal methods, one can show that if the true state $\hat\varrho$ 
is close to a generic MPS, then such a witness Hamiltonian always exists \cite{Perez-Garcia2007a}. What is more, it can be constructed from
the estimate of the algorithm sketched below. Its properties, chief
among them the gap, are efficiently computable.  In the Methods section,
we detail the efficient computation of these quantities, and we also consider how to treat states such as the GHZ for which local marginals alone are not quite sufficient for complete characterization (they violate our ``generic'' condition).

\subsection{An illustrative example}

We illustrate how our certification procedure operates if our estimate for
the state $\hat{\varrho}$ is a linear cluster state \cite{cluster}. The
cluster state is defined as the unique eigenstate (with eigenvalue $+1$)
of stabilizers $\hat{K}_i = \hat{Z}_{i-1} \hat{X}_i \hat{Z}_{i+1}$ for all $i=2,\ldots,N-1$ (together with boundary terms $\hat{X}_1 \hat{Z}_2$ and
$\hat{Z}_{N-1}\hat{X}_N$, which we do not treat separately for simplicity).
Assume that we have performed standard quantum state tomography on sets
of three neighbouring spins, $k=3$, which is the smallest useful set because
in a cluster state the rank of the reduced density matrices of contiguous blocks
is upper bounded by $R=4$. Let us now assume that on the basis of these data,
our procedure suggests that the linear cluster state is indeed our estimate.
The local Hamiltonian in this case is given by $\hat{H} =  \sum_i (\id-\hat{K}_i)/2$, where the $(\id-\hat{K}_i)/2=\hat{h}_i$ are projectors,
$\hat{H}$ has the cluster state as its unique ground state (with energy zero) and an energy gap $\Delta E=1$.  The fidelity between the cluster state $|\psi_{CS}\rangle$ and the state $\hat{\varrho}$ is bounded by $\langle \psi_{CS} | \hat{\varrho} |\psi_{CS}\rangle = 1 - \sum_i(\text{tr}[\hat{h}_i\hat{\sigma}_i] + \epsilon_i)$, where $\epsilon_i$ from Eq.~(\ref{eq:localerror}) quantifies the statistical and experimental error in the local experimental estimates, and $\text{tr}[\hat{h}_i\hat{\sigma}_i]$ quantifies how much the laboratory state $\hat{\varrho}$ deviates from an exact cluster state.

Tomography on a cluster state can also be performed with the scheme based 
on unitary transformations. A cluster state is the output of a quantum 
circuit where each qubit is initially prepared in the state $\ket{+}=1/\sqrt{2}\left(\ket{0}+\ket{1}\right)$ and a controlled phase transformation $CZ$ acts successfully on each pair of neighboring qubits. 
The $CZ$ gate changes the sign of state $\ket{11}$ and acts trivially on the 
other states of the computational basis. Thus, a cluster state is the output 
of a circuit whose structure corresponds to the one indicated on Fig.~\ref{fig_circuit} with $\kappa=2$ and the scheme based on unitary transformations directly applies.  Note that the unitary scheme takes advantage of the decreased rank of a reduced density matrix on the \emph{boundary} to save one qubit worth of local tomographic effort ($\kappa = 2$ vs.\ $k=3$).

\subsection{Scheme based on local measurements: Efficient determination of an MPS estimate}

With the experimentally obtained $\hat{\sigma}_i$, we now turn to the task
of finding an MPS $|\psi\rangle$ such that its reductions $\text{tr}_{1,\dots,i;i+k+1,\dots,N}[|\psi\rangle\langle\psi|]$ closely match
the $\hat{\sigma}_i$.
In other words: Let $\hat{P}_m^{i,k}$ be all possible products of Pauli
operators (enumerated by $m$) that act non-trivially only on sites 
$i+1,\dots,i+k$. Then, as the $\hat{P}_m^{i}$ are an orthogonal basis, 
the $\hat{\sigma}_i$ may be expanded as
\begin{equation}
    \hat{\sigma}_i = \frac1{2^N}\sum_m \text{tr}[\hat{\sigma}_i\hat{P}_m^{i}]\hat{P}_m^{i},
\end{equation}
where the expectations $\text{tr}[\hat{\sigma}\hat{P}_m^{i}]\in\rr$ are
obtained as results of tomographic measurements. Then we need to find a
matrix $|\psi\rangle\langle\psi|$ such that for all $m$ and $i$ the expectations $\text{tr}[|\psi\rangle\langle\psi|\hat{P}_{m}^{i}]$ coincide with those
of the tomographic estimates,
$\text{tr}[|\psi\rangle\langle\psi|\hat{P}_{m}^{i}] = \text{tr}[\hat{\sigma}_i\hat{P}_{m}^{i}]
$.

The method of choice for such a problem is singular value thresholding (SVT) \cite{matrix_completion,svthresh}, which has been developed very recently
in the context of classical {\em compressive sampling} or {\em matrix completion}
\cite{Compressed} and may also be applied to the quantum setting
\cite{Kosut, Gross,Gross2,Kosut2,Kosut3}. SVT is a recursive algorithm that 
provably converges to a low-rank matrix satisfying constraints of the type
$\text{tr}[|\psi\rangle\langle\psi|\hat{P}_{m}^i]=\text{tr}[\hat{\sigma}_i\hat{P}_{m}^i]
$. Unfortunately, a straightforward 
application of SVT requires time and memory that scale exponentially with the 
number of particles. However, a modification of the algorithm allows us to 
overcome this problem.

Given the measured values $\tr[\hat\sigma_i\hat P_{m}^{i}]$ the algorithm 
may then be described as follows. First set up the operator 
$\hat{R}=\sum_{m,i} \tr[\hat\sigma_i\hat P_{m}^{i}]\hat{P}_{m}^{i}/2^N$ and
initialize $\hat{Y}_0$ to some arbitrary matrix (e.g., the zero matrix). Then proceed recursively by finding the eigenstate $|y_n\rangle$ with largest 
eigenvalue, $y_n$, of $\hat{Y}_n$ and set
\begin{equation}
\begin{split}
\hat{X}_n&=y_n\sum_{m,i}\frac{\langle y_n|\hat{P}_{m}^{i}|y_n\rangle}{2^N}\hat{P}_{m}^{i},\\
\hat{Y}_{n+1}&=\hat{Y}_n+\delta_n(\hat{R}-\hat{X}_n).
\end{split}
\end{equation}
So far, this algorithm still suffers from the fact that in every step the
$2^N\times 2^N$ matrix $\hat{Y}_n$ needs to be diagonalized. However, the
$\hat{Y}_n$ are of the form $\sum_{m,i}^M a_{m,i}\hat{P}_{m}^{i}$, $a_{m,i}\in\rr$, where the $\hat{P}_{m}^{i}$ act non-trivially only
on sites $i+1,\dots,i+k$, i.e., they have the form of a local ``Hamiltonian". 
Hence, $|y_n\rangle$ can be determined as the highest energy state of this
Hamiltonian. For this task standard methods have been developed in condensed
matter physics \cite{Rommer 95,Schollwoeck} for which the number of parameters 
scale polynomially in the system size and converge rapidly \cite{Schuch,Aharonov} 
to the optimal MPS approximation.
The standard but exponentially inefficient SVT algorithm possesses a
convergence proof while our efficient modification does not. Hence, we 
now present numerical examples for different target states $|\phi\rangle$ 
to demonstrate the feasibility and efficiency of the proposed algorithm.

These numerical examples suggest convergence of our algorithm to a MPS that
closely matches the experimentally obtained reductions $\hat{\sigma}_i$.
To arrive at the fidelity bound we follow the steps of Section~\ref{S:certification}:
(i) Obtain estimates $\hat{\sigma}_i$ of the reductions to $k$ adjacent spins $\hat{\varrho}_i$ of the state in the laboratory such that
$\|\hat{\sigma}_i-\hat{\varrho}_i\|_{\text{tr}}\le\epsilon_i$, (ii) compute the expectations $p_{m_i}=\text{tr}[\hat{\sigma}_i\hat{P}_{m}^i]$,
which are the input to the MPS-SVT algorithm,
(iii) obtain a MPS estimate $|y_n\rangle$, the reductions of which closely match the $\hat{\sigma}_i$ by utilizing the MPS-SVT algorithm.
As $|y_n\rangle$ is an MPS, one can then efficiently obtain a \textit{parent Hamiltonian} \cite{Perez-Garcia2007a} and a
lower bound, $\Delta$, on the energy gap above the ground state (see Methods for details).
Putting all this together, the above programme returns a state $|y_n\rangle$,
its parent Hamiltonian $\hat{H}=\sum_i\hat{h}_i$, and a number $\Delta$ such that
\begin{equation}
\langle y_n|\hat{\varrho}|y_n\rangle \ge
 1-\frac{\sum_i(\epsilon_i+\text{tr}[\hat{h}_i\hat{\sigma}_i])}{\Delta}.
\end{equation}
\begin{figure}
\includegraphics[width=0.9\columnwidth]{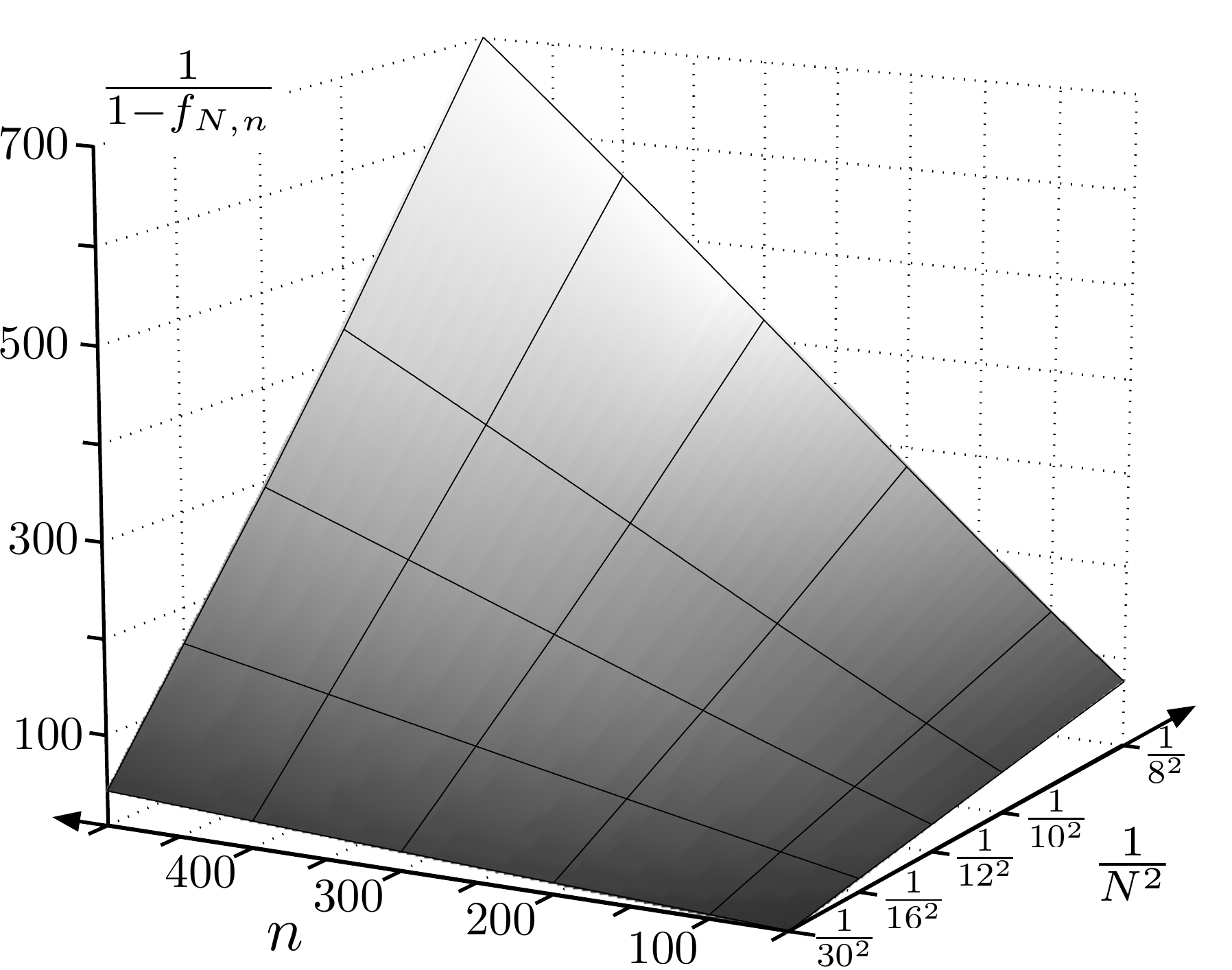}
\caption{\label{fig_ising} Performance of the MPS-SVT algorithm for the ground state (the target state $|\phi\rangle$) of the critical Ising model. We chose $k=2$, i.e., only nearest neighbour reductions are used to reconstruct the state.  Plot illustrates the scaling of the error in the fidelity $1-f_{N,n}=1-|\langle \phi|y_n\rangle|^2\sim N^2/n$ with the number of spins $N$ and the iterations $n$ of the algorithm.}
\end{figure}

\subsection{Example: Strongly interacting quantum systems}

We start with ground states of nearest-neighbor Hamiltonians on a chain, i.e., 
the $|\phi\rangle=|gs\rangle$ are completely determined by all the reductions to 
two adjacent spins and we expect the above algorithm not only to produce states 
that match the reduced density matrices of the ground states but, in fact,
states that are themselves close to the ground states. Among ground states of one-dimensional nearest-neighbor Hamiltonians, those at a critical point are 
the most challenging to approximate by MPS as they violate the entanglement 
area law \cite{EisertCP}.  We demonstrate the effectiveness of 
our algorithm for such an example: the critical Ising model  in a transverse field on a chain of length $N$ with open boundary conditions. The 
ground state of this Hamiltonian is unique and hence it is completely 
characterized by its reductions to $k=2$ neighbouring sites. In other words, if 
we find an MPS that has the same reductions, the fidelity will be one.
We proceed as follows. (i) We obtain the reductions $\hat{\varrho}_i$ to two
neighbouring sites of the true ground state, (ii) from these reductions we 
obtain the expectations
$p_{m_i}=\text{tr}[\hat{\varrho}_i\hat{P}_{m_i}]$, which are the input to the MPS-SVT algorithm. In Fig.~\ref{fig_ising} we show the fidelity 
$f_{N,n}=|\langle y_n|\phi\rangle|^2$ of the true ground state $|\phi\rangle$ 
and the $n$'th iterate of the above algorithm as a function of $n$ and the 
length $N$ of the chain. It shows that for fixed system size $1-f_{N,n}$ 
decreases as $\sim\! 1/n$ while for fixed $n$ it increases slower than $N^2$.
This provides an indication that our algorithm is polynomial in the system size.
\begin{figure}
\includegraphics[width=\columnwidth]{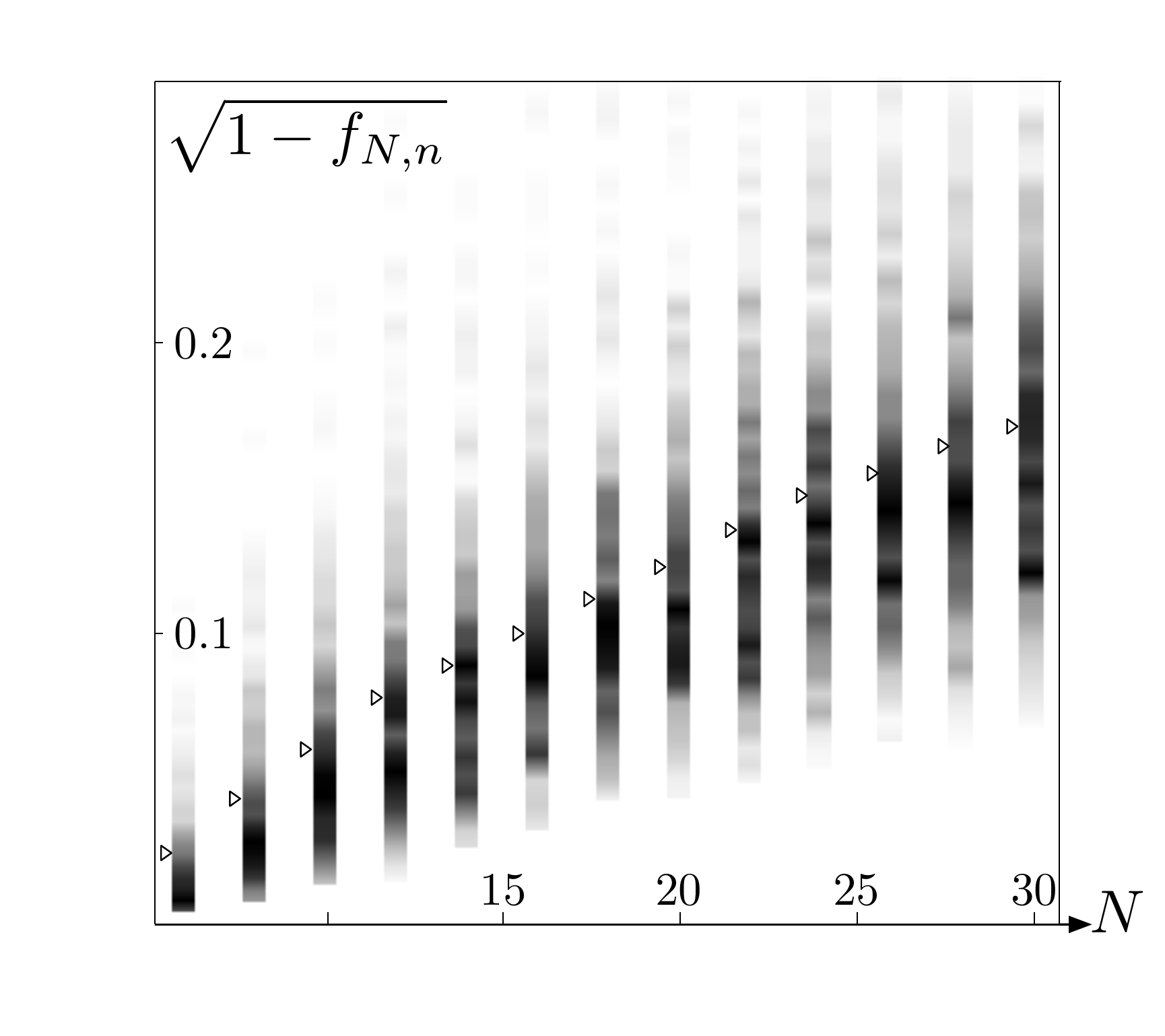}
\caption{\label{fig_random} Fidelity $f_{N,n}=|\langle \phi|y_n\rangle|^2$, for a fixed value $n=5$, as a function of the number of spins $N$ for the ground state (the target state $|\phi\rangle$) of the random nearest-neighbour Hamiltonians of Eq.~\ref{E:randomH}. As in Fig.~\ref{fig_ising}, we chose $k=2$ to reconstruct the state. The plot shows densities  (arrows indicate the mean) obtained from 1000 random realizations. Similar to the Ising model, the scaling for fixed $n$ of $1-f_{N,n}$ is better than $\sim \!N^2$.
}
\end{figure}

The Ising model is solvable and, in order to show that we are not considering 
a special case that is particularly favourable, we also 
consider one-dimensional random Hamiltonians of the form
\begin{equation}\label{E:randomH}
 \hat{H}=\sum_{i=1}^{N-1}\hat{r}^{(i)}_i\hat{r}^{(i)}_{i+1},
\end{equation}
where the $\hat{r}^{(i)}_i$, $\hat{r}^{(i)}_{i+1}$ act on spin
$i$ and $i+1$, respectively, and are hermitian matrices with
entries that have real and imaginary part picked from a uniform
distribution over $[-1,1]$. For each Hamiltonian, as before, we first
determine the ground state $|gs\rangle$ (our target state
$|\phi\rangle$) and its reductions and then computed
the fidelity $f_{N,n}$ after $n$ iterations
of the MPS-SVT algorithm, see Fig.~\ref{fig_random}.

\subsection{Example: W-state preparation in ion traps}

\begin{figure}
\includegraphics[width=0.9\columnwidth]{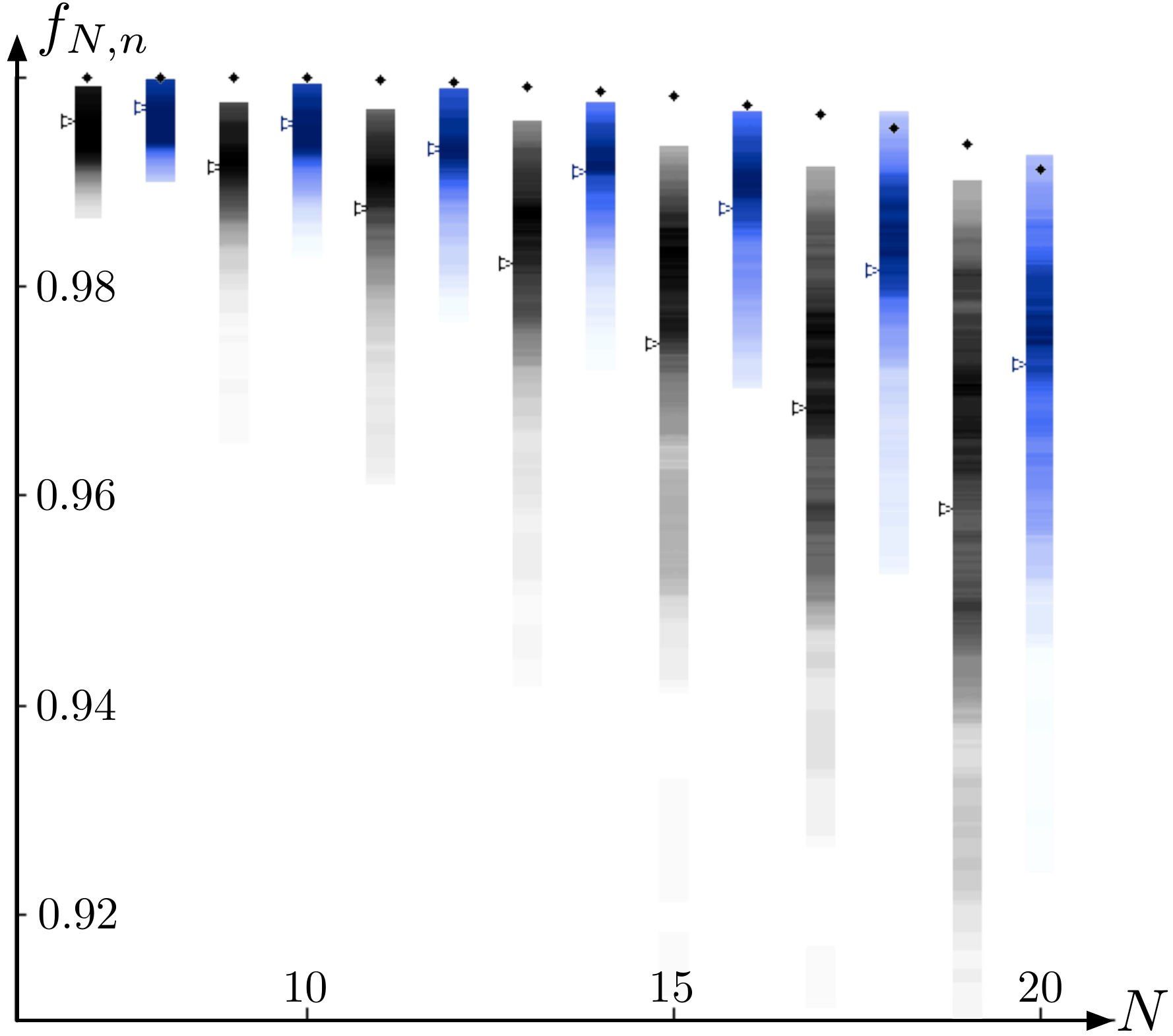}
\caption{\label{fig_w} Performance of the MPS-SVT algorithm for W-states,
$|\phi\rangle=(|10\cdots 0\rangle+|010\cdots 0\rangle+\cdots+|0\cdots 01\rangle)/\sqrt{N}$.
We are not limited to few ions and demonstrate
convergence for up to $20$ ions -- even higher number of ions are easily
accessible due to the MPS alteration of
the SVT method, demonstrating the efficiency of our approach. We mimic experimental noise by adding Gaussian
distributed random numbers with zero mean to the local expectations $p_{m_i}=\text{tr}[\hat{\sigma}_i\hat{P}_{m_i}]$ and show results for $n=4000$ MPS-SVT iterations. Plot shows $f_{N,n}=|\langle \phi|y_n\rangle|^2$ as a function of the number of ions, $N$, for no noise (dots) and Gaussian noise (densities obtained from 100 realizations for each $N$, arrows indicate mean) with a standard deviation of $0.005$ (blue; only even values of $N$ plotted for clarity) and $0.01$ (black, odd $N$).}
\end{figure}

Our method is of interest for many situations in which standard tomography will not be feasible. This
is the case for the verification of state preparation in
experiments with too many particles. An example is the recent ion trap experiment
\cite{Haffner 05} for the preparation of W-states,
$|\phi\rangle=(|10\cdots 0\rangle+|010\cdots 0\rangle+\cdots+|0\cdots 01\rangle)/\sqrt{N}$,
that were limited
to 8 qubits principally because the classical postprocessing of
data became prohibitive for longer chains. Here we demonstrate
the efficiency of our approach (we are not limited to few ions and demonstrate
convergence for up to $20$ ions -- even higher number of ions are easily
accessible due to the MPS alteration of
the SVT method) by illustrating how one would
postprocess experimentally obtained reduced density matrices
to guarantee the generation of $|\phi\rangle$ or a state very
close to it. Note that, as in the above spin chain examples, one need only take tomographic data on pairs ($k=2$) of neighbouring qubits.  We mimic experimental noise by adding Gaussian
distributed random numbers with zero mean to the $p_{m_i}$. After
initializing the algorithm with $\hat{Y}_0=\hat{R}$, where we obtain $\hat{R}$
from the MPS representation of
$|\phi\rangle$, we use 
$x_n:=\sum_{m_i}|p_{m,i}-\langle y_n|\hat{P}_{m}^{i,k}|y_n\rangle|$ as a figure
of merit for convergence, i.e., after a given number of iterations,
we pick the $|y_n\rangle$ with minimum $x_n$. The result of such
a procedure is shown in Fig.~\ref{fig_w}.

\section{Discussion}

We have presented two schemes that efficiently produce an MPS description as a tomographic estimate of a quantum state, along with a tight fidelity bound.  We emphasize that no assumptions are necessary for our scheme; if no such MPS description exists, this will be evident from the local tomographic data and our schemes will herald a failure.  However, the enormous successes of MPS for describing both one-dimensional quantum systems found in nature as well as a host of states relevant to quantum information ensures that our methods will be very useful in practice.

So far we presented the method for pure states
and one-dimensional systems. Various generalizations are possible: Our first scheme using unitary control can also handle mixed states corresponding to small ensembles of pure MPS. Suppose we are presented with a state $\hat{\varrho}$ that is a mixture of $M$ pure states, each of which is an MPS having bond dimension $D$. Then the reduction of $\hat{\varrho}$ to any subsystem has rank at most $MD$. We can proceed as before, performing unitary operations on blocks of $\kappa = \lceil\log_d(MD)\rceil + 1$ sites, in order to disentangle individual sites from the rest of the chain. At the end of the chain, we will find a mixed state $\eta$ of rank $M$ on the last $\kappa-1$ sites. We decompose $\eta$ as a mixture of $M$ pure states. This yields a representation of $\hat{\varrho}$ as an ensemble of $M$ pure MPS, each with bond dimension at most $dMD$.

Our second scheme using local measurements can also be extended to handle mixed states. While the $k$-particle reduced
density matrices do not uniquely determine the mixed state $\hat{\varrho}$,
reconstructions of better and better quality can be obtained
by increasing $k$. As an example, suppose $\hat{\varrho}$ is the Gibbs
state corresponding to a $k$-local Hamiltonian $\hat{H}$, i.e. the
state minimizing the free energy
\begin{equation}
\text{tr}[\hat{\varrho}\hat{H}]-TS(\hat{\varrho}).
\end{equation}
The first term is, as before, determined by the reduced density matrices. The entropy of
the total state however can only be learned exactly from the complete density matrix.
Yet, for essentially all reasonable physical systems, the entropy density
$\lim_{k\rightarrow\infty} S(\tr_{k+1,...}(\hat{\varrho}))/k$
in the thermal state of a Hamiltonian exists \cite{BratelliRobinson} (In particular, finitely correlated states are precisely those states whose entropy density is exactly equal to $ S(\tr_{k+1,...}(\hat{\varrho})) -  S(\tr_{k,...}(\hat{\varrho}))$ for some finite value of $k$).
As a consequence, the total entropy of the state can be
estimated efficiently from knowledge of the reduced density matrices.
Hence, our second scheme using local measurements may be extended to mixed states by
considering purifications and can then, e.g.,  also handle thermal states of local Hamiltonians, under the physically reasonable assumption that the entropy density exists \cite{BratelliRobinson}. In addition, it can be generalized to all mixed finitely correlated states (FCS) \cite{FannesNW 92}, though it is not always possible to certify the resulting estimates.

Higher-dimensional systems are more challenging, because the most straightforward generalization of MPS, known as projected entangled-pair states (PEPS) \cite{peps}, cannot be computed as efficiently. However, the certification method using a frustration-free parent Hamiltonian remains efficient in the case of qubits with nearest-neighbor couplings~\cite{Beaudrap2010}. Combinations of our techniques can be used to reconstruct other classes of states, such as tree tensor networks \cite{ttn} and multiscale entanglement renormalization ansatz (MERA) states \cite{mera}, for which efficient heuristics for minimizing local Hamiltonians are available.

\acknowledgements
The authors acknowledge discussion with F.G.S.L. Brand{\~a}o at early
stages of this project. The work at Ulm University has been supported
by the EU Integrated Project QAP, the EU STREP's CORNER and HIP and an
Alexander von Humboldt Professorship. SDB acknowledges the support of the
Australian Research Council and the Perimeter Institute.
STF and RS were supported by the Perimeter Institute
for Theoretical Physics. Research at Perimeter is
supported by the Government of Canada through Industry
Canada and by the Province of Ontario through the Ministry
of Research \& Innovation. DG is glad to acknowledge
support from the EU (CORNER). OLC and DP are partially funded by NSERC and FQRNT. YKL is funded by the US ARO/NSA.

\section{Methods}
We start by recalling the MPS representation of a state $|\psi\rangle$ with open boundary conditions (generalisations to periodic boundary conditions are entirely straightforward).
\begin{equation}
\label{MPS}
|\psi\rangle=\sum_{s_1=0}^{d_1-1}\cdots\sum_{s_N=0}^{d_N-1}M_1[s_1]\cdots M_N[s_N]|s_1\cdots s_N\rangle,
\end{equation}
where the $M_i[s]$ are $D_i\times D_{i+1}$ matrices with $D_1=D_{N+1}=1$. We denote the \textit{bond dimension} by $D=\max D_i$.

\subsection{Direct tomography}

This method proceeds by disentangling all the qudits of the chain sequentially. Thus, it  yields a valid MPS description if
every unitary exactly disentangles one qudit. Put another way, while it is crucial to obtain a good estimate
of the $d^{\kappa-1}$-dimensional subspace on which $\hat{\sigma}$ is
supported, it is \emph{not necessary} to identify the eigenvectors
of $\hat{\sigma}$ exactly : \emph{any} set of orthonormal vectors generating the subspace
is sufficient for our tomography procedure and leads to an MPS description
in another gauge \cite{Perez-Garcia2007a}. This property will be central to our error analysis.

To understand the effect of errors and imperfections in our tomography procedure, consider the very first step of the recursive procedure. Tomography is performed on the first $\kappa$ sites to ideally find a state with non-maximal support, and unitary $\hat{U}_1$ is applied to rotate that state into the subspace $\mathcal{H}_1^{\textrm{cutoff}} = \ket 0\otimes  (\mathbb{C}^d)^{\tensor n-1}$. In any experimental setting, the resulting state $\hat{U}_1 \ket\phi$ would surely not lie entirely in that subspace. This can be either because the state of the system is
not exactly an MPS with bond dimension $D$, but merely close
to one or because our estimate of the density matrix $\hat{\sigma}$ on sites
$1$ to $\kappa$ is slightly wrong due to measurements that are,
in practice, noisy and restricted to finite precision. Indeed, we can expect that the reduced density matrix on the first $\kappa$ sites will actually
be full-rank, though most of its probability mass will lie on a subspace
of dimension at most $D$. So, each time we apply a disentangling operation $\hat{U}_{i}$,
we also want to \textit{truncate} the reduced state to the desired subspace. Similarly, a faulty estimate of $\hat{\sigma}$ will result in a small probability mass that lies outside the estimated support.

Given an estimated disentangling unitary $\hat{U}_1$, any state $\ket{\phi}$ can be expressed as
\begin{equation}
\hat{U}_1\ket{\phi} = \frac{\ket 0 \otimes \ket{\eta_1} + \ket{e_1}}{\sqrt{1+\braket{e_1}{e_1}}}
\label{eq:step1}
\end{equation}
where $\ket{e_1}$ is some sub-normalized error vector supported on the subspace orthogonal to $\mathcal{H}_1^{\textrm{cutoff}}$. The unitary $\hat{U}_1$ is chosen to minimize the norm $\epsilon_1 = \braket{e_1}{e_1}$ of this error vector. It is possible to directly estimate the magnitude of the error $\epsilon_1$ by measuring the first qudit in the standard basis; the error is equal to the population of the non-zero states.

In subsequent steps of the recursion, we are given a state of the form
\begin{equation}
\hat{U}_{i}\ldots \hat{U}_1\ket\phi = \frac{\ket 0^{\otimes i} \otimes \ket{\eta_i} + \ket{e_i^{\textrm{cm}}}}{\sqrt{1+\braket{e_i^{\textrm{cm}}}{e_i^{\textrm{cm}}}}}
\end{equation}
where $\ket{e_i^{\textrm{cm}}}$ is the accumulated error vector that lies in the subspace orthogonal to $\mathcal{H}_i^{\textrm{cutoff}} = \ket 0^{\otimes i}\otimes  (\mathbb{C}^d)^{\tensor N-i}$. As a first step, we can truncate this error vector by measuring the first $i$ particles in the standard basis and post-select on the all-zero outcome. This occurs with a probability roughly $1-\epsilon_i^{\textrm{cm}}$, and leaves the system in the state $\ket 0^{\otimes i} \otimes \ket{\eta_i}$. We then repeat the steps leading to Eq.~\ref{eq:step1} with the post-selected state $\ket{\eta_i}$. The resulting state will be
\begin{align}
\hat{U}_{i+1}\hat{U}_i\ldots \hat{U}_1\ket\phi & =  \frac{\ket 0^{\otimes i} \otimes \hat{U}_{i+1}\ket{\eta_i} + \hat{U}_{i+1}\ket{e_i^{\textrm{cm}}}}{\sqrt{1+\braket{e^{\textrm{cm}}_i}{e^{\textrm{cm}}_i}}} \\
& =  \frac{\ket 0^{\otimes i+1} \otimes\ket{\eta_{i+1}} + \ket{e_{i+1}^{\textrm{cm}}}}{\sqrt{1+\braket{e_{i+1}^{\textrm{cm}}}{e_{i+1}^{\textrm{cm}}}}}
\end{align}
where
\begin{equation}
\ket{e_{i+1}^{\textrm{cm}}} = \frac{\ket{e_i} + U_{i+1}\ket{e_i^{\textrm{cm}}}}{\sqrt{1+\braket{e^{\textrm{cm}}_i}{e^{\textrm{cm}}_i}}}
\end{equation}
and therefore $\epsilon^{\textrm{cm}}_{i+1} \leq \epsilon^{\textrm{cm}}_{i} + \epsilon_{i+1}$.

Thus, we see that errors accumulate {\em linearly} with the number of particles; if we denote $\ket{\psi} = \hat{U}^{\dagger}_1\ldots \hat{U}^{\dagger}_{N-\kappa+1} \ket 0^{N-\kappa+1}\ket{\eta_{N-\kappa+1}}$ the estimated MPS, we have
\begin{equation}
\left\Vert \ket{\phi}-\ket{\psi}\right\Vert =\left\Vert \ket{e_{N-\kappa+1}^{\textrm{cm}}}\right\Vert \leq\sum_{i=1}^{N-\kappa+1}\left\Vert \ket{e_{i}}\right\Vert \leq N\,\epsilon
\end{equation}
where $\epsilon=\max_{i}\left\Vert \ket{e_{i}}\right\Vert $.
The overall error is at most the sum of the individual errors on each
step.  In addition, each of the $\epsilon_i$ is revealed during the tomographic procedure because they correspond to the post-selection success probability. This provides a direct method to {\em certify} the inferred state.

\subsection{Parent Hamiltonians}
Let $|\psi\rangle$ be as in (\ref{MPS}) and such that $\sum_sM_i[s]M_i[s]^\dagger=\id$ for all $i=1,\dots,N$. This can always be achieved by subsuming qudits at the beginning and end of the chain into qudits with higher dimension and successive singular value decompositions \cite{FannesNW 92, Perez-Garcia2007a}.
Now let $N,k\in\nn$ such that $ N/k\in\nn$, and assume that the MPS is injective \cite{Perez-Garcia2007a} such that for all $j=k,\dots,N-2k$ the set
\begin{equation}
\left\{M_{j+1}[s_1]\cdots M_{j+k}[s_k]\,\big|\,s_i=1,\dots,d_i\right\}
\end{equation}
spans $\cc^{D_{j+1}\times D_{j+k+1}}$. Then $|\psi\rangle$ is the unique ground state of
\begin{equation}
\hat{H}=\sum_{n=0}^{N/k-2}\hat{P}_n,
\end{equation}
where $\hat{P}_n$ is the projector onto the subspace orthogonal to the range of the mapping $\Gamma_n:\cc^{D_{nk+2k+1}\times D_{nk+1}}\rightarrow \cc^{d_{nk+1}\cdots d_{nk+2k}}$,
\begin{equation}
\begin{split}
\Gamma_n(X)=\!\!\!\sum_{\substack{s_{nk+1},\\ \dots,\\ s_{nk+2k}}}\!\!\!\text{tr}\big[X M_{nk+1}[s_{nk+1}]\cdots M_{nk+2k}[s_{nk+2k}]\big]\\
\times|s_{nk+1}\cdots s_{nk+2k}\rangle.
\end{split}
\end{equation}
To get an efficiently computable lower bound the energy gap, we use
\begin{equation}
\Delta E=\max\left\{\lambda\,\big|\,\hat{H}(\hat{H}-\lambda)\ge 0\right\}.
\end{equation}
We find
\begin{equation}
\hat{H}^2=\hat{H}+\sum_{\substack{n,m\\ n\ne m}}\hat{P}_n\hat{P}_m\ge
\hat{H}+\!\!\!\!\!\!\sum_{\substack{n,m\\ 1\le |n- m|k\le 2k}}\!\!\!\!\!\!\frac{\hat{P}_n\hat{P}_m+\hat{P}_m\hat{P}_n}{2},
\end{equation}
where we omitted non-negative summands. Now consider the following quantity, which will bound the individual terms in the previous equation,
\begin{align}
\gamma_{n,m}&=\min\left\{\lambda\,\big|\,\hat{P}_n\hat{P}_m+\hat{P}_m\hat{P}_n+\lambda(\hat{P}_n+\hat{P}_m)\ge 0\right\}\nonumber\\
&=\min\left\{\lambda\,\big|\,(\hat{P}_n+\hat{P}_m)^2\ge (1-\lambda)(\hat{P}_n+\hat{P}_m)\right\}\nonumber \\
&=1-\max\left\{\lambda\,\big|\,(\hat{P}_n+\hat{P}_m)^2\ge \lambda(\hat{P}_n+\hat{P}_m)\right\},
\end{align}
where the maximum is given by the smallest non-zero eigenvalue of $\hat{P}_n+\hat{P}_m$. Hence
\begin{equation}
\hat{H}^2\ge \hat{H}-\!\!\!\!\!\!\sum_{\substack{n,m\\ 1\le |n- m|k\le 2k}}\!\!\!\!\!\!\gamma_{n,m}\hat{P}_n
\ge (1-\gamma)\hat{H},
\end{equation}
and therefore we have the lower bound $\Delta E\ge 1-\gamma$, where
\begin{equation}
\gamma=\max_{0\le n\le N}\sum_{\substack{m\\ 1\le |n- m|k\le 2k}}\!\!\!\!\!\!\gamma_{n,m}.
\end{equation}

Injectivity may fail to hold in certain singular cases. The simplest
example given by the family of GHZ-type states $\frac1{\sqrt
2}(\ket{0, \ldots, 0} + e^{i\phi} \ket{1,\ldots, 1})$. Since any local
reduced density matrix is independent of $\phi$, it is
\emph{impossible} to distinguish the members of that family based on
local information alone. Indeed, in this example, the ground state
space of $\hat H$ will be two-dimensional, spanned by $\ket{0,\ldots,
0}$ and $\ket{1,\ldots, 1}$. One may check that the gap analysis above
continues to hold in the degenerate case (unlike the original in
\cite{FannesNW 92}), now certifying the overlap between $\hat\varrho$
and the ground-state space. This fact alone implies an exponential
reduction in the number of unknown parameters. It is easy to see that
the small remaining uncertainty about $\hat\varrho$ can be lifted in
our example by measuring the ``string operator'' $\sigma_x
\otimes\ldots\otimes \sigma_x$, which has expectation value
$\cos\phi$. Using e.g.\ the results of \cite{walgate}, the method of
this particular example immediately generalizes to any MPS with
a two-fold degeneracy, such as the W state. Higher degeneracies may be treated by
straight-forward, but more tedious, methods.

\end{document}